# Femtosecond laser-assisted selective holding with ultra-low power for direct manipulation of biological specimens

Krishangi Krishna,[1] Joshua A. Burrow,[1] Zhaowei Jiang,[2] Wenyu Liu,[1] Anita Shukla,[2] and Kimani C. Toussaint Jr. [1,3,*]

[1]*PROBE Lab, School of Engineering, Brown University, Providence, Rhode Island 02912, USA*
[2]*School of Engineering, Brown University, Providence, Rhode Island 02912, USA*
[3]*Brown-Lifespan Center for Digital Health, Providence, Rhode Island 02912, USA*
*\*kimani_toussaint@brown.edu*

**Abstract:** Traditional optical tweezers techniques often rely on high-power continuous wave (CW) lasers, which can introduce unwanted thermal effects and photodamage to delicate samples. To overcome these limitations, we demonstrate femtosecond laser assisted selective holding with ultra-low power (FLASH-UP). We find that the FLASH-UP exhibits a five times greater trap stiffness than CW-OT, and can trap at lower intensities. Furthermore, we demonstrate OT of different pathogenic bacteria species and find that FLASH-UP does not impact cell motility. These results pave the way for applications in sorting, bio-sensing, in vivo cell manipulation and single cell analysis.

## 1. Introduction

Optical tweezers (OT), originally conceptualized by Arthur Ashkin in the 1970s [1-3], remains a valuable, non-invasive tool for cellular manipulation ranging from individual to populations of cells in aqueous environments [4-6]. Specifically, OT has been used for cell sorting [7], live cell patterning [8], single cell optoporation and transfection [9], and characterizing the mechanical properties of cells in different growth media [10], collagen [11], and DNA [12]. OT is a result of balancing the electromagnetic gradient and scattering forces in a trap, a phenomenon that is a consequence of light's radiation pressure. The net force acting on a particle caught in an optical trap is on the order of piconewtons, which is on the same scale as forces produced by the naturally occurring mechanoenzymes in biology [4,13,14]. A typical OT setup is constructed from an inverted, laser-scanning optical microscope, whereby a single-wavelength, continuous-wave (CW) laser source delivering powers in the range of 5 to 100 mW (corresponding to intensities of ~22 – 77.5 MW/cm$^2$) for strong, diffraction-limited focusing is employed [15-18]. For specific applications for biological manipulation, optical wavelengths in the near-infrared regime and intensities on the order of MW/cm$^2$ are used [4,6,19].

The laser-induced heating resulting from light absorption in the cell cytosol causes thermal damage to the trapped biological specimens and the generation of toxic reactive oxygen species [14,20,21]. Researchers have explored various methods to mitigate this issue such as reducing the total heat generated by trapping at $T = 0°C$ [22, 23], minimizing the duration that trapped cells are exposed to the laser [14,21], and relying on indirect trapping methods that involve tweezing dielectric particles that are tethered to the biological species of interest [10,24,25]. Nonetheless, these alternate methods limit the ability to directly interrogate cell-cell interactions. In the specific case of trapping bacteria, it has been shown that the local heating results in adverse effects that impact their propagation, motility, and expression of stress-response genes, even when exposed to the minimal threshold intensities required for trapping

[20,26]. Importantly, these traditional cases of OT for biological manipulation use CW lasers. An alternative approach has stemmed from the use of ultrafast, femtosecond (fs)-pulsed lasers operating at MHz-repetition rates, where the high peak powers and short pulse durations result in nJ pulse energies that are less deleterious to cells [27,28].

In the Rayleigh regime (for particles of diameter $d <$ wavelength of light $\lambda$), fs-OT has indirectly trapped DNA by being tethered to a micron sized dielectric [27]. The merits of employing fs-OT over one that is CW have been exemplified for this regime where fs-OT has trapped particles with average powers an order of magnitude lower than that required for CW. Due to the particle size, there is an increase in thermal diffusion which is offset by the ultra-high peak powers of pulsed lasers with the generation of transient forces [29,30]. In the Lorentz-Mie ($d \approx \lambda$) and Mie ($d > \lambda$) regimes, fs-OT has directly trapped bacteria such as *E. coli* and eukaryotic cells such as *Phaffia rhodozyma* and red blood cells, respectively. However, the power-densities used to trap these cells has led to cell membrane dysfunction and formation of damaged white spots after being trapped for 1-15 minutes [26,28,31]. Notably, these investigations have been conducted using input laser powers on the order of a few mW, with little to no exploration of OT efficiency within the sub – 1 mW regime.

In this work, we demonstrate femtosecond laser-assisted selective holding with ultra-low power (FLASH-UP) to achieve optical tweezing using average powers deep in the µW for objects in the Lorentz-Mie regime. We apply FLASH-UP to pathogenic bacterial cells with pili including *Staphylococcus aureus*, *Bacillus paranthracis*, *Staphyloccocus epidermidis*, and *Vibrio cholerae*. Lastly, we tweeze *Pseudomonas aeruginosa*, an infectious bacteria species with a single polar flagellum often causing pneumonia or leading to bacterimia. We find that FLASH-UP demonstrates a trap stiffness 5× greater than CW-OT for dielectric spheres, and can trap at power densities half that of CW. To a first approximation, the use of an ultrafast source introduces a transient force component that works in concert with the traditional gradient force to achieve stable optical trapping. The implications of these results will allow for the characterization of bacteria on a single-cell level offering insights into mechanistic clues hidden in bulk measurements and a stronger understanding of pathogenic mechanisms [32].

## 2. Theory

It is generally complicated to model femtosecond OT in the Lorentz-Mie regime, but valuable insights can be obtained from examining effects in the Rayleigh model. The time-varying electric field E of a Gaussian beam propagating along the *z*-axis and polarized along $\hat{x}$ is expressed as [33]

$$\boldsymbol{E}(\rho, z, t) = \hat{x} \frac{iE_0}{i + 2z/kw_0^2} exp\left[i\omega_0 t - ikz - \frac{i2kz\rho^2}{(kw_0^2)^2 + 4z^2} - \frac{kw_0^2 \rho^2}{(kw_0^2)^2 + 4z^2}\right] exp\left[-\frac{\left(t - \frac{z}{c}\right)^2}{\tau^2}\right], \quad (1)$$

where $w_0$ is the beam waist at z = 0, and $\rho, k = \frac{2\pi}{\lambda}, \omega_0, t, \tau$ are the radial coordinate, wave number, carrier frequency, time, and pulse duration, respectively. The resultant Lorentzian force $\boldsymbol{F_p}$ arising from spatial inhomogeneous field distributions acting on a Rayleigh particle is given by

$$\boldsymbol{F_p}(\rho, z, t) = [\boldsymbol{p}(\rho, z, t) \cdot \boldsymbol{\nabla} \boldsymbol{E}(\rho, z, t)] + [\partial_t \boldsymbol{p}(\rho, z, t)] \times \boldsymbol{B}(\rho, z, t) = \boldsymbol{F}_{grad} + \boldsymbol{F}_t \quad (2)$$

where $\boldsymbol{p} = \alpha \boldsymbol{E}$ and $\boldsymbol{B}$ are the dipole moment and corresponding magnetic flux. Here, $\alpha = 4\pi n_m^2 \varepsilon_0 r^3 \left[\frac{m^2 - 1}{m^2 - 1}\right]$, and $\varepsilon_0, r, m$ are vacuum dielectric permittivity, radius of particle, and ratio of the refractive indices of particle $n_p$ to medium $n_m$, respectively. The forces generated can thus be decomposed into

$$\boldsymbol{F}_{grad,\rho} = \frac{\xi \tilde{p}}{(1+4\tilde{z}^2)}, \tag{3}$$

$$\boldsymbol{F}_{grad,z} = \frac{2\xi \tilde{z}}{kw_0}\left[\frac{1+4\tilde{z}^2-2\tilde{p}^2}{(1+4\tilde{z}^2)^2}\right], \text{ and} \tag{4}$$

$$\boldsymbol{F}_t = \left[\frac{\tilde{z}kw_0^2}{c\tau} - \tilde{t}\right]\left[\frac{\xi w_0}{c\tau} - \frac{8\mu_0 I(\rho,z,t)}{\tau}\right], \tag{5}$$

where $(\tilde{p}, \tilde{z}, \tilde{t}) = (\rho/w_0, z/w_0, t/\tau)$ and $I(\rho, z, t)$ are the normalized temporal spatial coordinates and intensity of electric field, and $\xi = \frac{2\alpha I(\rho,z,t)}{n_m \varepsilon_0 c w_0}$. Thus, the use of an ultrafast optical pulse introduces an auxiliary force term given in Eq. 5, which would otherwise be zero when using a traditional CW source.

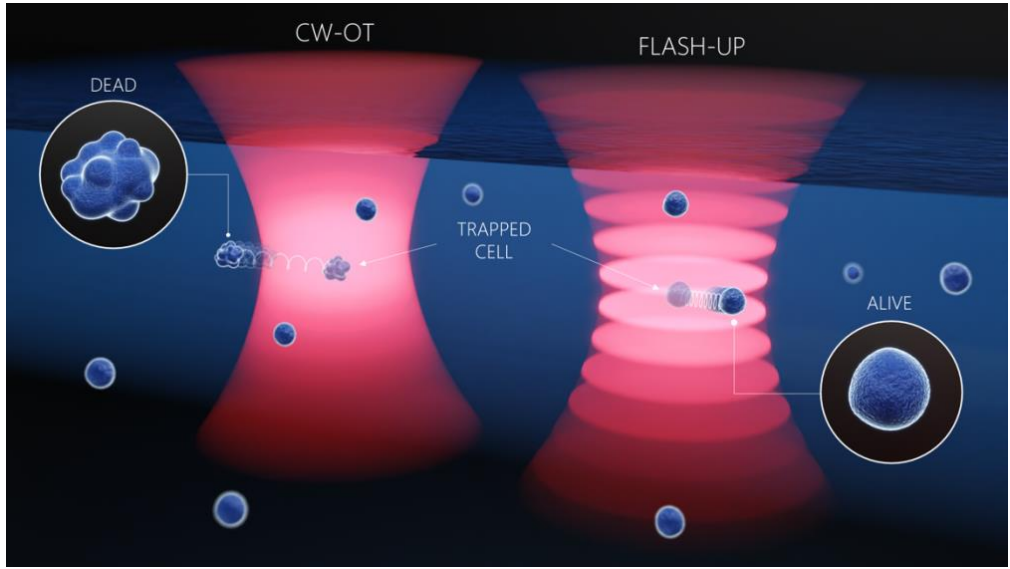

Fig 1. Concept art highlighting optical trapping using CW and FLASH-UP. Conventional CW (left) and femtosecond (right) lasers optically trapping cells in an aqueous environment. A cell after being trapped with the CW laser has undergone deformations, whereas a cell trapped by FLASH-UP retains its morphology and experiences a stronger optical force.

We observe that the temporal force depends inversely on the pulse duration. Lastly, the scattering force, which pushes the particle along the axial direction, is given by $\boldsymbol{F}_{scat} = \hat{z}\frac{\sigma n_m}{c}I(\rho,z,t)$. Here, $\sigma = \frac{8\pi}{3}(kr)^4 r^2 \left[\frac{m^2-1}{m^2-1}\right]$ is the particle cross section.

## 3. Experimental setup

As a proof-of-concept demonstration, we first employ a conventional single-beam OT experimental setup, shown in Fig. 2, and compare the trap stiffness on 1-µm diameter silica microspheres using a 120 femtosecond-pulsed (Spectra Physics, Insight X3) or a continuous wave (Newport, LQC905-85E) laser source operating at a central wavelength of 905 nm, which can be accessed by a flip mirror. We employ a customized Olympus IX83 inverted microscope.

The laser source is spatially filtered and expanded to produce a TEM$_{00}$ Gaussian intensity profile. It passes through the linear polarizer, and neutral density (ND) filter for power adjustment at the sample plane. The laser is directed onto a 2D galvanometer that controls the position of the laser. The laser is guided into the microscope by a dichroic mirror (OCT-21020-BX3TRF) and focused to a spot size of ~425 nm by a 40X/1.3 NA oil immersion objective lens (Olympus). The condenser lens (Nikon) collects the forward scattered light and redirects it into the beam splitter such that the light can be focused into the quadrant photodiode (QPD; Thorlabs, PD80A) by a lens. The $x$ and $y$ voltage outputs from the QPD are recorded using an oscilloscope (Picoscope, 3203D) from which the trap stiffness, κ, is extrapolated via the power spectrum method. Visualization of the sample is enabled by a 150 W halogen fiber optical light source (Amscope, HL150-AY) and time-lapse sequences are imaged with an sCMOS (Hamamatsu, C15440-20UP).

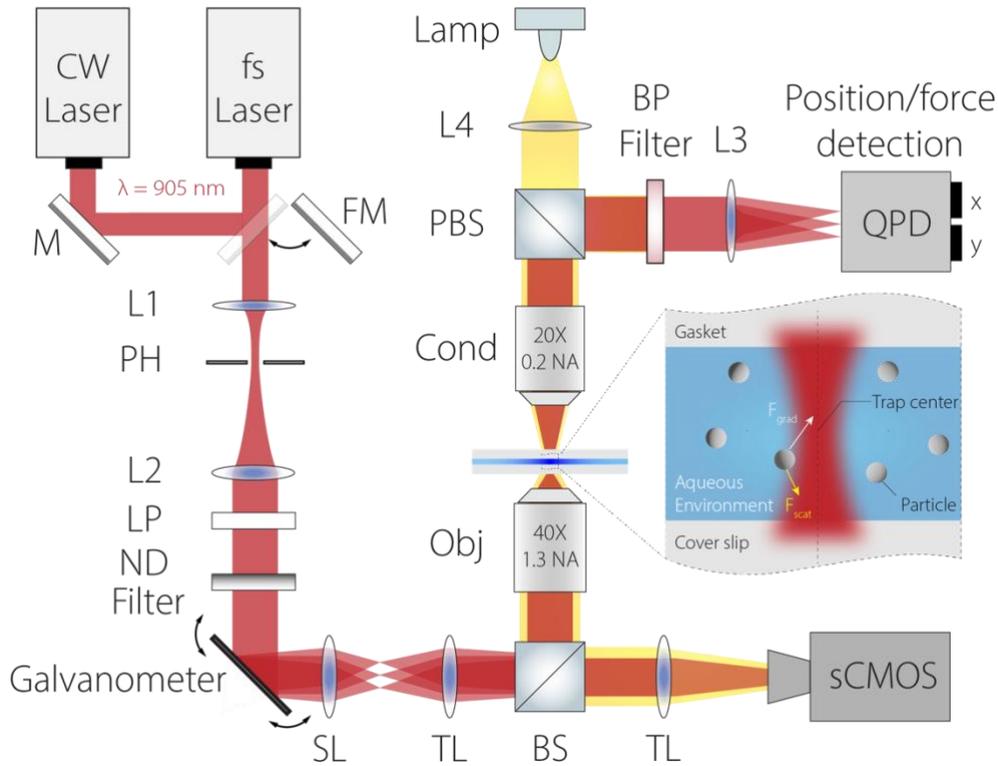

Fig 2. A CW or pulsed laser source are employed for optical trapping of particles. L: lens, PH: pinhole, LP: linear polarizer, ND: neutral density, SL: scanning lens, TL: tube lens, BS: beam splitter, PBS: polarizing beam splitter, BP: bandpass filter, QPD: quadrant photodiode.

## 4. Results

### 4.1 Observation of ultra-low power optical trapping of dielectric spheres

Output voltages are recorded from a QPD when a particle is trapped. Minimal trapping thresholds $I_{thr}$ and trapping efficiencies $Q = Fc/nP,$ where $F$, $c$, $n$ and $P$ are the force exerted on the particle, speed of light, particle refractive index, and average power, respectively, are determined for a range of average intensities from 1.41 mW/μm² ($P = 80$ μW) – 17.67 mW/μm² ($P = 1$ mW). Figure 3(a) depicts a trapped microsphere (magenta frame) with the femtosecond laser source at time $t = 0$ s. When the trap is turned off at $t = 20.87$ s, the particle is released from the trap and is no longer within the trapping plane (green). However, when the laser is reactivated at $t = 20.87$ s, the particle returns to its initial trapped position (magenta) at $t = 21.37$ s, and remains trapped for the remainder of the experiment ($t = 35$ s). The corresponding voltage displacement collected by the QPD is seen in Fig. 3(b). The particle is confined to a harmonic potential well [Fig. 3(c)] and the κ is rendered by fitting the power spectrum to a single-sided Lorentzian [Fig. 3(d)].

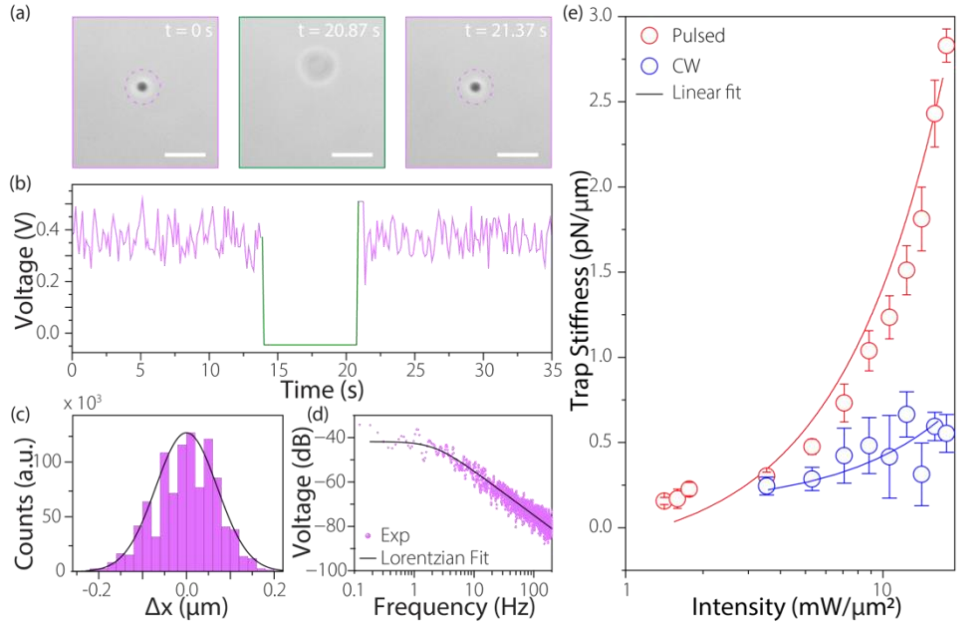

Fig. 3. (a) Time-lapse sequence of a silica microsphere in (magenta) and out (green) of a trap when the laser is on/off. Dotted circles represent laser position. (b) Corresponding voltage output vs. time relating to particle position. (c) 1D position histogram of particle within the trap. (d) Fitting of single-sided Lorentzian to extrapolate the corner frequency. (e) Trap stiffness dependence on input intensity for pulsed (red) and CW (blue) lasers. Data is presented as mean values +/- standard deviation of mean. Scale bars, 5 μm.

Figure 3(e) depicts the trap stiffness of FLASH-UP (red) and CW (blue) lasers for the microspheres. The experiment is conducted on 3 different particles and values are given as mean ± standard deviation (SD). The values of $I_{thr}$ for FLASH-UP and CW lasers are 1.41 and 3.51 mW/μm$^2$, respectively. Trap stiffness κ ranges from 0.30 – 2.82 pN/μm for pulsed, and 0.25 – 0.55 pN/μm for CW. To quantitatively evaluate the trap strength, we apply a fit (slope = $\eta$) that matches well with experimental data. We find that FLASH-UP not only has $\eta = 0.14$ pN·μm·mW$^{-1}$ while for CW $\eta = 0.025$ pN·μm·mW$^{-1}$, but also stably traps a single particle at a lower intensity threshold.

### *4.2 Trapping bacterial cells*

To investigate the use of FLASH-UP for biological applications, we perform OT experiments on bacterial cells. We selected four differently shaped pathogenic bacteria, namely *S. aureus* (spherical), *B. paranthracis* (rod), *V. cholerae* (jellybean), and *S. epidermidis* (elliptical). Morphological characterization of each cell is given in Fig S1. in Supplemental 1. The average aspect ratios and areas of *S. aureus* are 1.21 and 0.40 μm², *B. paranthracis* are 3.01 and 2.04 μm², *V. cholerae* are 2.28 and 1.10 μm², and *S. epidermidis* are 1.26 and 0.37 μm², respectively.

Trap stiffnesses are determined similarly as described in the previous section. Figure 4 (a-d) displays the trap stiffness of each cell type for both pulsed (red) and CW (blue) lasers. The insets of each graph are scanning electron microscopy images obtained for each bacteria type. We find that FLASH-UP demonstrates a stronger trap stiffness for each cell type when compared to using a CW source. The $I_{thr}$ for each cell is shown in Fig. 3(e). Moreover, the $\eta$ shown in Fig. 3(f) depicts that the trap stiffness using FLASH-UP consistently exhibits a higher stiffness than CW regardless of bacteria-type. Interestingly, this intensity may go down even lower depending on the trapping media used due to differences in refractive indices and thermal properties changing the trapping conditions [34]. In Fig. S2 in Supplemental 1 we measured the trap stiffness of silica microspheres in deionized (DI) water and 1× phosphate buffer saline (PBS) of both laser sources. We observed that the trap stiffness of the dielectrics for both lasers in 1× PBS were lower than in DI water at intensities above 1.76 mW/μm² for FLASH-UP and 7.06 mW/μm² for CW.

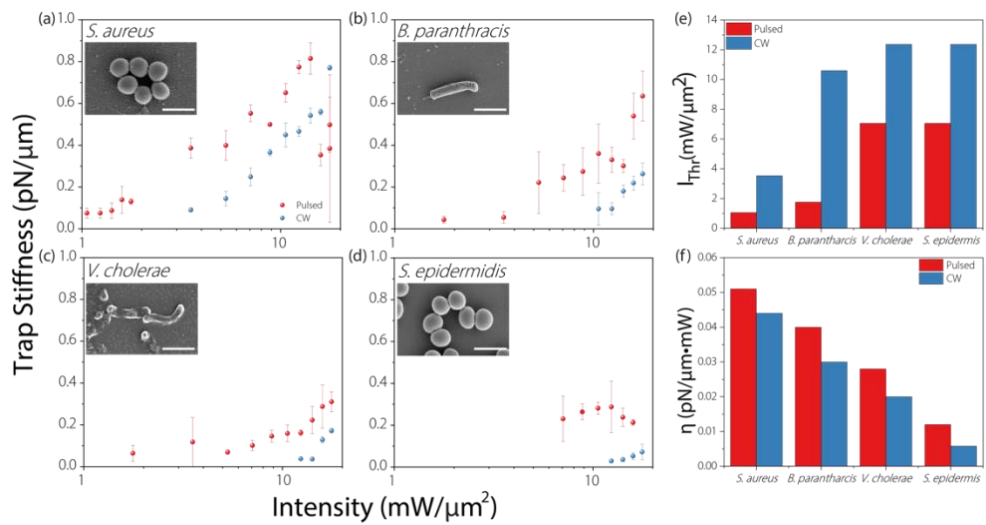

Fig. 4. Comparing trap stiffness (mean ± SD) for pulsed (red) and CW (blue) lasers based on intensity for (a) *S. aureus* (b) *B. paranthracis* (c) *S. epidermidis* (d) *V. cholerae*. (e) Threshold intensities required for trapping each cell. (f) Trap efficiencies for each cell. Scale bars, 5 μm.

To understand the optothermal effects on cell viability, image sequences are captured at a time resolution of 5 μs and cell position is assessed with ImageJ. Interestingly, we note that *S. aureus* cells are immobilized after they were briefly trapped by the CW laser. However, this is not observed with FLASH-UP. Fig S3. in Supplemental 1 plots the position of *S. aureus* for 60 s after the cell was trapped for 90 s with either FLASH-UP (red) or CW (blue) laser sources at intensities of 8.83 mW/μm². It is evident that there is minimal movement of the cell once it has been trapped by the CW laser source as the SD of cell position is 10× lower than FLASH-UP.

Figure 5(a) depicts the trap stiffness of *P. aeruginosa* for FLASH-UP (red) and CW (blue) lasers. Morphology characterization is given in Fig S2. in Supplemant 1, and the average aspect ratio and area are 2.30 and 0.64 μm². The $I_{thr}$ for trapping these cells are 1.76 mW/μm² and 7.06 mW/µm² for the pulsed and CW sources, respectively. FLASH-UP also exhibits a stiffer trap for all intensities when compared to the CW. We then plot the trajectory of the cells along the *x-y* plane with respect to time post-trap for 5 s for pulsed (top) and CW (bottom) sources Fig. 4(b).

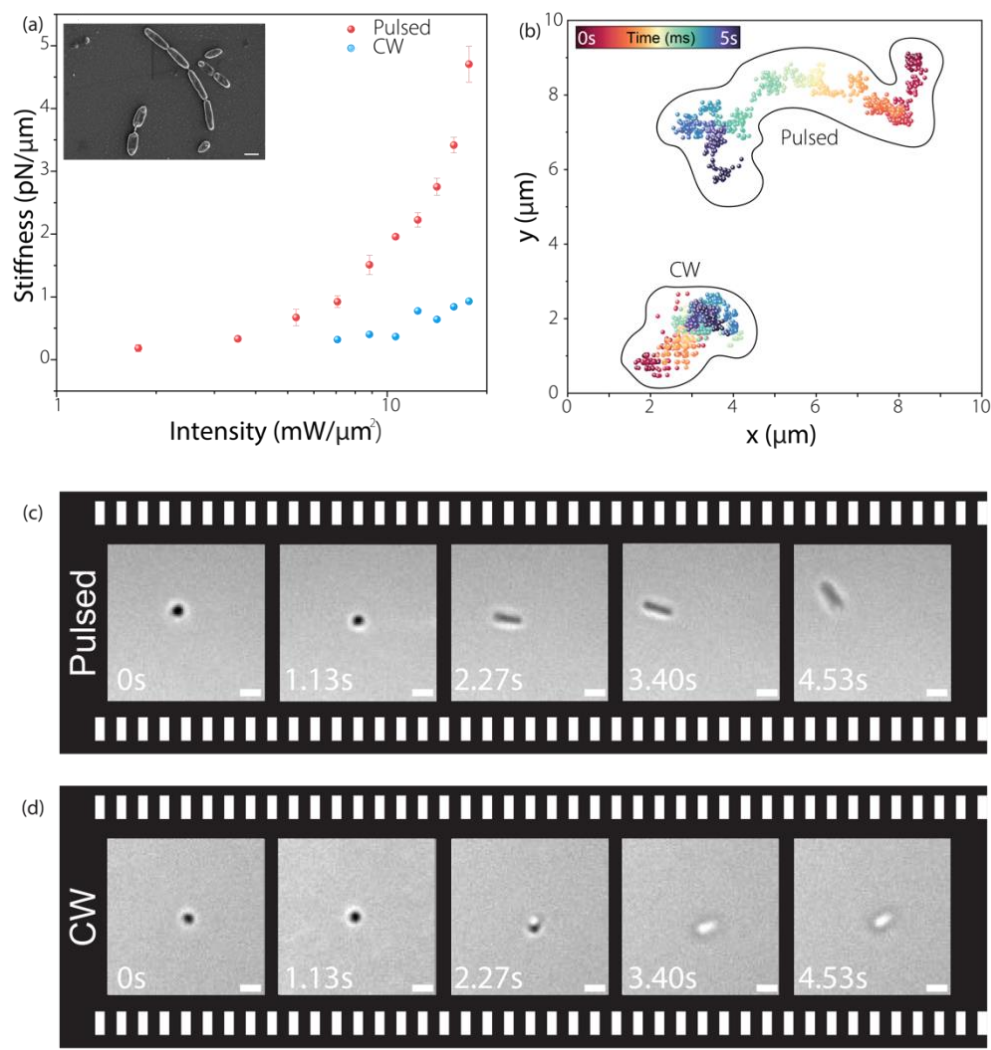

Fig 5. (a) Comparing trap stiffness (mean ± SD) of pulsed (red) and CW (blue) based on intensity for *P. aeruginosa*. (b) Cell trajectory with respect to time after being trapped for 5 s. Time-lapse sequence of a cell once the trap is turned off for (c) pulsed and (d) CW. Scale bars, 1 µm.

## 5. Discussion

In summary, we investigated and compared the trap stiffness for CW-OT and FLASH-UP on dielectric microspheres within the Lorentz-Mie regime for average powers below 1 mW. To our knowledge, this is the first demonstration of ultra-low power trapping with a fs source at intensities as low as 1.41 mW/µm². We consistently observed that FLASH-UP exhibited 5× the trap stiffness compared to the CW, and $I_{thr}$ for the pulsed trapping was half that for CW trapping with 3.51 mW/μm$^2$. Subsequently, we successfully trapped pathogenic bacterial cells including *S. aureus, B. paranthracis, S. epidermidis, V. cholera*, and *P. aeruginosa* using intensities as low as 1.76 mW/μm$^2$ with FLASH-UP. Furthermore, we observed a reduction in the mobility of *S. aureus* and *P. aeruginosa* when trapped with the CW laser but not with FLASH-UP, suggesting that the CW source may affect cell functionality and motility at these low power densities. The superior effectiveness of fs-OT over CW-OT is attributed to the generation of a transient force that is inversely related to the pulse duration, and these results reveal important differences that could be leveraged for the direct manipulation of other biomolecules.

Enhanced cell retention over extended durations opens avenues for investigating dynamic biological processes such as gene expression and protein localization during cell division. This capability extends to seamlessly transporting cells across diverse channels with varying compositions, and conducting diagnostic mechanical testing at the cellular level. Specifically, leveraging the potential of FLASH-UP for ultra-low-power trapping of bacteria holds promise for in-depth exploration of microbial adhesion, bacterial surface colonization, and the formation of biofilms. Employing OT allows for the precise manipulation of bacterial spacing, facilitating a comprehensive understanding of how cell density influences cell-to-cell signaling. Furthermore, aligning bacteria with each other provides valuable insights into their preferred growth directions. FLASH-UP potentially could be used to deepen comprehension of biofilm formation, a common occurrence on medical implants, surgical fixations, and vascular replacements, but also advance our knowledge in manipulating cellular behavior for biomedical applications.

Future work to improve the effectiveness of FLASH-UP includes optimizing the trap stiffness as a function of pulse width. Additionally, incorporating cell-viability assays help establish the maximum duration cells can be trapped with FLASH-UP without compromising cell morphology and functionality.

**Funding.** Brown University.

**Acknowledgments.** Special thanks to Geoff Williams from the Brown University Leduc Bioimaging Core Facility for his invaluable assistance with SEM. The SEM was procured through a high-end instrumentation grant from the Office of the Director at the National Institutes of Health (S10OD023461). We thank Dr. Jay Tang (Brown University) for helpful discussions regarding cell trapping. We thank Collin Polucha (Brown University) for careful editing of the research paper. J.A.B. was supported by the Hibbitt Post-Doctoral Fellowship in the Brown School of Engineering and the Postdoctoral Diversity Enrichment Program of the Burroughs Welcome Fund.

**Disclosures.** The authors declare no conflicts of interest.

**Data availability.** All data needed to evaluate the conclusions in the paper are present in the paper and/or the Supplementary Materials. Additional data related to this paper may be requested from the authors.

**Supplemental document.** See Supplmental for supporting content.

**FEMTOSECOND LASER ASSISTED SELECTIVE HOLDING WITH ULTRA-LOW POWER FOR DIRECT MANIPULATION OF BIOLOGICAL SPECIMENS: SUPPLEMENTAL DOCUMENT**

## 1. Materials and Methods

*1.1 Sample preparation*

In this study, silica microspheres and biological samples were prepared. Silica microspheres (Thermofisher; n~ 1.4) were obtained and diluted with DI water in a 1:1000 ratio. *S. epidermidis* ATCC12228, *S. aureus* ATCC25923, *B. paranthracis* 13061, and *V. cholerae* N16961 were purchased from American Type Culture Collection (ATCC). *P. aeruginosa* PA01 were acquired from Walter Reed Army Medical Center. Sterile Lysogeny Broth (LB) (Sigma Alderich, St. Louis, MO) was prepared according to the manufacturer's instructions. LB consisted of 10 g/L tryptone 10g/L Sodium chloride (NaCl) 5 g/L yeast extract. The strain was maintained in 25% glycerol (v/v) in LB at -80ºC until use. A sterile inoculating loop was used to streak the bacterial strain onto a nutrient broth agar plate. The culture was inoculated using a single colony and incubated at 37ºC with 200 rpm orbital shaking for 3 hours. All bacteria cultures were diluted to optical density at 600 nm ($OD_{600}$) of 0.01 and washed twice via centrifugation with 1× phosphate-buffered saline (1× PBS) at 4,000 g for 10 minutes. The cell pellet was resuspended in 1× PBS. Here, 20 µL of either the silica spheres or cells was dropcasted onto a gasket (Thermofisher) and sandwiched with a 0.17-mm thick coverslip to prevent dehydration of the sample, and mounted with the coverslip side facing the objective lens.

*1.2 Scanning electron microscope (SEM) imaging*

Overnight cultures of each bacterial strain were prepared. Bacteria were inoculated from the overnight culture onto sterilized silicon wafers (0.5 cm × 0.5 cm) in 24-well plates and incubated at 37°C for 16-18 hours. The silicon wafers were washed twice with 1× PBS and fixed using 1× PBS supplemented with 2% (v/v) glutaraldehyde and 2% (v/v) paraformaldehyde for 4 hours at 4°C. Samples were then dehydrated using a graded ethanol series from 50% (v/v) to 100% (v/v), with each gradient lasting for 10 minutes, and then allowed to dry at room temperature for 24 hours. These samples were sputter-coated with gold-palladium for 2.5 minutes at 20 mA under argon. Samples were examined using an environmental SEM (Quattro S, ThermoFisher Scientific, Waltham, MA) operated at 2 kV.

*1.3 Trap Stiffness*

An optical trap can be modeled by an overdamped harmonic oscillator described at room temperature as [35]

$$\dot{x}(t) + 2\pi f_c x(t) = \sqrt{2D}\Omega(t), \tag{1}$$

where *m* is the mass of the particle, $f_c = \frac{\kappa}{2\pi\beta_0}$ is the roll-off frequency, $\beta_0$ is the Stokes drag coefficient, $\Omega(t)$ and represents the random perturbations due to Brownian motion. The associated power spectrum $P(f)$ describing the particle fluctuations is determined by taking the Fourier transform of equation (1) which is representative of a Lorentzian as

$$P(f) = \frac{k_B T}{\pi^2 \beta_0 (f^2 + f_c^2)}. \tag{2}$$

We extract $f_c$ by fitting the experimental data, and to determine $\kappa = 2\pi\beta_0 f_c$.

To determine trap stiffness, we record the particle displacements within a 5 s time frame at a sampling rate of 100 kHz using both laser sources. This is repeated 3 times at each power to ensure reliability. The focal power density is calculated with $I_0 = \frac{P_0}{A}$, where $P_0$ is the input optical power measured at the sample plane after the objective lens, and the focal-spot area is $A = \pi w_0^2$. Here, $w_0 = \frac{0.61\lambda}{NA}$, where $\lambda$ and NA are the free space input wavelength and numerical aperture of the objective lens, respectively.

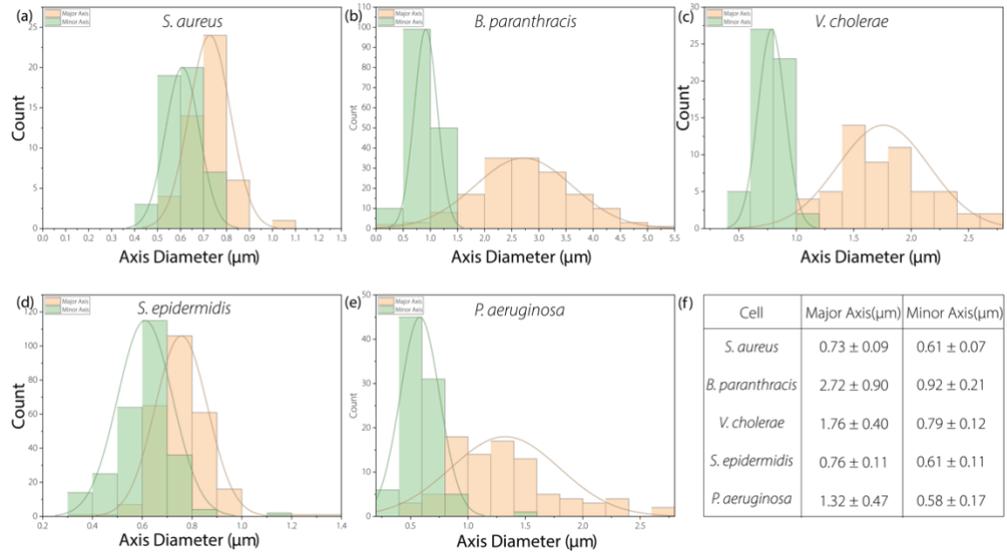

Fig. S1. Morphology histograms of (A) *S. aureus*, (B) *B. paranthracis*, (C) *V. cholerae*, (D) *S. epidermidis*, (E) *P. aeruginosa*. (F) Major and minor axes values of each cell.

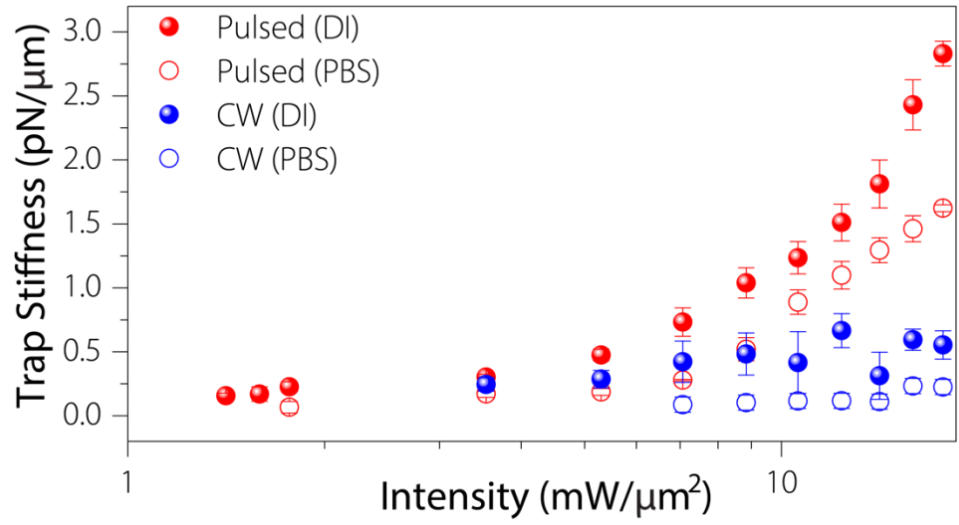

Fig. S2. Trap stiffness (mean ±SD) of silica microspheres vs. intensity for the pulsed (red) and CW (blue) in deionized (DI) water and 1× phosphate buffer saline (PBS).

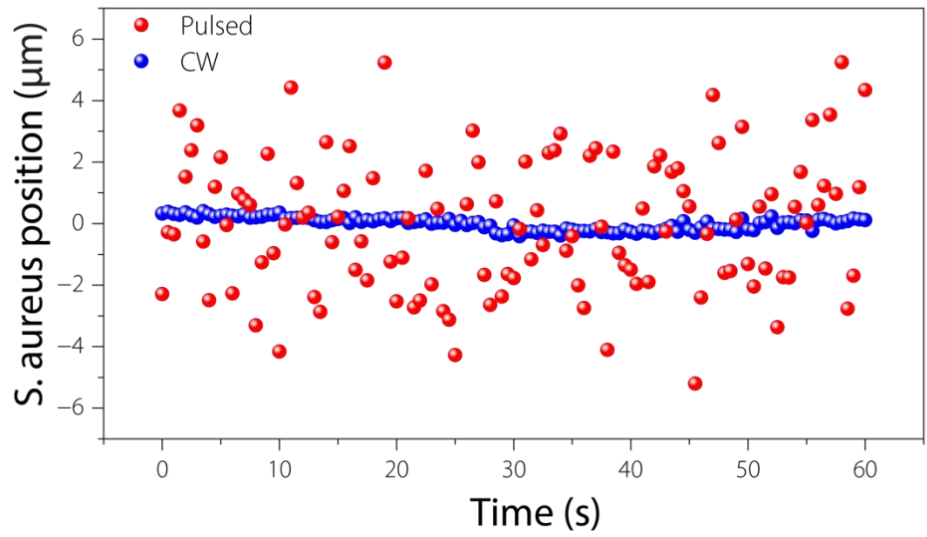

Fig S3. Position of *S. aureus* after it has been trapped for 90s at 8.83 mW/μm$^2$ by pulsed (red) and CW (blue) with respect to time.